This version of the article has been accepted for publication after peer review but is not the Version of Record and does not reflect post-acceptance improvements, or any corrections. The Version of Record is available online at: https://doi.org/10.1038/s41467-026-78072-0.

# Unexpected stabilization of a single-*q* charge density wave in pristine 1*T*-$TiSe_2$

Kimoon Higashihira Han,[1,†] Dongju Hwang,[1,†] Dongsin Kim,[2,†] Yeojin Ahn,[1] Jaehun Cha,[1,3] Seonggeon Gim,[1] Gyubin Lee,[1,4] Mingi Jho,[1] Chan-young Lim,[1,5] Sae Hee Ryu,[6,7] Chris Jozwiak,[6] Aaron Bostwick,[6] Eli Rotenberg,[6] Jonathan D. Denlinger,[6] Makoto Hashimoto,[8] Donghui Lu,[8] Young Woo Choi, [2,*] Gil Young Cho,[1,9,*] and Yeongkwan Kim[1,*]

[1]*Department of Physics, Korea Advanced Institute of Science and Technology, Daejeon 34141, Republic of Korea*

[2]*Department of Physics and Center for Nano Materials, Sogang University, Seoul 04107, Republic of Korea*

[3]*Department of Physics, Pusan National University, Busan 46241, Republic of Korea*

[4]*Center for Semiconductor Technology, Korea Institute of Science and Technology, Seoul 02792, Republic of Korea*

[5]*Donostia International Physics Center, 20018 San Sebastián, Spain*

[6]*Advanced Light Source, Lawrence Berkeley National Laboratory, Berkeley, California 94720, USA*

[7]*Korea Basic Science Institute, Daejeon 34133, Republic of Korea*

[8]*Stanford Synchrotron Radiation Lightsource, SLAC National Accelerator Laboratory, Menlo Park, California 94025, USA*

[9]*Center for Artificial Low Dimensional Electronic Systems, Institute for Basic Science, Pohang 37673, Republic of Korea*

[†]These authors contributed equally.

*Corresponding authors. E-mail: ywchoi02@sogang.ac.kr; gilyoungcho@kaist.ac.kr; yeongkwan@kaist.ac.kr

**Over the past decade, studies revisiting the classical charge-density-wave system 1*T*-$TiSe_2$, long understood to host the 2 × 2 triple-*q* order with equivalent subcomponents, have suggested that these subcomponents may not be identical, challenging a 50-year-old paradigm. Here, we demonstrate not only that these subcomponents are non-identical but also that they develop sequentially rather than simultaneously. By tracking folded-band features, angle-resolved photoemission spectroscopy reveals that a unidirectional 2 × 1 single-*q* order stabilizes first, while the other two components emerge sequentially at lower temperatures, completing the triple-*q* phase. Our free-energy analysis explains this hierarchy through repulsive coupling among the three subcomponents, as supported by first-principles calculations incorporating phonon anharmonicity. Consequently, the three subcomponents are inequivalent in both phase and amplitude. These results reconcile reported symmetry breaking and differing characteristic temperatures and establish a framework in which ostensibly simple, high-symmetry charge-density order emerges from temperature-dependent competition and cooperation among inequivalent subcomponents.**

Charge density waves (CDWs) are often perceived as conventional phenomena, owing to their seemingly simple concept and frequent occurrence in low-dimensional materials.[1,2] In two or higher dimensions, most CDWs are unconventional: their microscopic origins remain elusive,[3] the standard Peierls mechanism often fails to apply,[4-6] and their phase diagrams raise persistent questions about how CDW order competes or intertwines with superconductivity, magnetism, and other electronic instabilities.[7-9] Over the past decade, experimental advances have further enriched this complexity, revealing additional symmetry breakings such as chirality, nematicity, and reduced rotational symmetry that call for a reassessment of this long-studied phenomenon.[10-12]

A paradigmatic system encapsulating many of these issues is the layered transition-metal dichalcogenide 1$T$-$TiSe_2$. Its CDW transition at $T \approx 200$ K cannot be explained by simple Fermi surface nesting,[2,13] prompting a long-standing debate about its origin, with proposals involving excitonic interactions,[14-19] electron-phonon coupling,[20-25] or a combination of both.[26-28] Cu intercalation between the layers suppresses the CDW and induces superconductivity near the quantum critical point of the CDW.[29] Moreover, scanning tunnelling microscopy, x-ray scattering, and optical measurements have reported evidence of further broken symmetries below $T_{CDW}$,[30-33] renewing interest in a material intensively studied since the 1970s.

Here, we push the frontier further. We show that the threefold CDW in 1$T$-$TiSe_2$ does not form in a single step at $T_{CDW}$. Instead, only one of the three symmetry-related components—a unidirectional 2 × 1 modulation (single-$q$ CDW)—condenses first at $T_{CDW}$, with the remaining two components stabilizing sequentially at lower temperatures. Landau free energy analysis reveals that this sequence is driven by repulsive, rather than attractive, interactions between the three CDW order parameters. This hierarchy resolves, in a coherent manner, debates over earlier reports of inequivalent $q$-vectors and associated symmetry lowering, including chirality, [30-33] and calls for a revised framework for understanding CDWs in general, in which unidirectional subcomponents play a central role.

## Results and Discussion

The high-temperature normal state of 1$T$-$TiSe_2$ crystallizes in the space group $P\bar{3}m1$, a simple hexagonal lattice with threefold rotational symmetry. Its low-energy electronic structure consists of a Se 4$p$ hole band centred at the $\bar{\Gamma}$ point and a Ti 3$d$ electron band at the $\bar{M}$ point of the Brillouin zone (Fig. 1b). The triple-$q$ CDW phase,[34] the widely believed charge-order structure of 1$T$-$TiSe_2$, gives rise to a 2 × 2 × 2 superlattice, which folds the Se hole band at $\bar{\Gamma}$ to the three symmetry-related $\bar{M}$ points—$\bar{M}_1$, $\bar{M}_2$, and $\bar{M}_3$—and vice versa (Fig. 1b). In angle-resolved photoemission spectroscopy (ARPES), the resulting folded bands display spectral weight comparable to that of the original band.[5,14,35]

In the presence of intact threefold rotational symmetry, the folded bands at the three $\bar{\mathrm{M}}$ points must be equivalent. Therefore, any deviation from this equivalence signals a breaking of the threefold symmetry. To capture possible deviations directly, we set up an experimental geometry, aligning the geometrical mirror plane defined by the light incidence plane to one of the mirror planes of the crystal, $\bar{\Gamma}$-$\bar{\mathrm{M}}_2$ direction (Fig. 1c). The analyser slit was oriented perpendicular to the mirror planes, enabling simultaneous measurement of the folded bands across the mirror plane at $\bar{\mathrm{M}}_1$ and $\bar{\mathrm{M}}_3$. The incident light polarization was chosen to be symmetric with respect to the aligned mirror planes, ensuring that the photoemission matrix elements do not introduce extrinsic asymmetries in the signal strength from $\bar{\mathrm{M}}_1$ and $\bar{\mathrm{M}}_3$. This configuration provides a direct, symmetry-respecting comparison of the initial states at these two $\bar{\mathrm{M}}$ points, making it possible to detect potential discrepancies between them.

Fig. 1d shows temperature-dependent momentum-distribution curves (MDCs) at a binding energy of 0.35 eV, revealing folded bands at the $\bar{\mathrm{M}}_1$ and $\bar{\mathrm{M}}_3$ points. The intensity at $\bar{\mathrm{M}}_1$ is systematically stronger than at $\bar{\mathrm{M}}_3$, providing clear evidence of broken symmetry between the two $\bar{\mathrm{M}}$ points. More importantly, the two folded features display distinct temperature dependences. Across the known CDW transition temperature at $T_{\mathrm{CDW}} = 200$ K, the $\bar{\mathrm{M}}_1$ spectral weight broadens and weakens, signalling the loss of long-range order. By contrast, the $\bar{\mathrm{M}}_3$ spectral weight remains diffuse well below $T_{\mathrm{CDW}}$ and becomes discernible only at much lower temperatures, around 100–120 K.

These differences are not a trivial consequence of an overall intensity change, nor of extrinsic artefacts, including matrix element effects (Supplementary Note 1), as evident in representative energy–momentum cuts at three temperatures (Fig. 1e–g). At $T = 245$ K ($> T_{\mathrm{CDW}}$), folded features at both $\bar{\mathrm{M}}_1$ and $\bar{\mathrm{M}}_3$ are weak and broad, confined to a narrow energy window near the Fermi level, consistent with short-range CDW fluctuations above $T_{CDW}$[14,36-38] (Fig. 1e). At $T$ = 175 K (< $T_{CDW}$), $\bar{M}_1$ exhibits well-defined spin-split upper and lower valence branches whose tails extend to higher binding energies, whereas $\bar{M}_3$ remains weak and diffuse without a clear dispersion (Fig. 1f). Only below $T$ = 100 K does a sharp dispersive feature emerge at $\bar{M}_3$ (Fig. 1g). This behaviour implies that, at $T_{CDW}$, only the component of the CDW order parameter that folds the $\bar{\Gamma}$ hole band to $\bar{M}_1$ (the $q_1$ component) condenses, while the components associated with $q_2$ and $q_3$ remain short-ranged until much lower temperatures.

To test this interpretation, we mapped the electronic structure over the entire surface Brillouin zone at several temperatures. Constant-energy maps at $E = E_F - 0.35$ eV are shown in Fig. 2a; curvature-processed images from the same dataset are provided in Fig. 2b to enhance feature visibility.[39] Above $T_{CDW}$, short-range fluctuations produce broad, nearly featureless ellipses at the $\bar{M}$ points. At $T$ = 200 K, sharp concentric rings—arising from two folded branches derived from the upper and lower hole bands—appear only at $\bar{M}_1$, consistent with folding by the $q_1$ vector of a 2 × 1 ordering, while contours connected by $q_2$ and $q_3$ remain diffuse. Upon cooling to 167 K, the $\bar{M}_2$ points connected by $q_2$ begin to show comparable sharp rings. By $T$ = 100 K, all $\bar{M}$ points are equally sharp. These observations demonstrate a sequential stabilization of the three 2 × 1 CDW components: long-range order is first established for $q_1$, then for $q_2$, and finally for $q_3$, as summarized schematically in Fig. 2c.

Next, to determine the characteristic temperatures of the sequential transitions, we quantified the temperature dependence of the folded-band spectral weight for the $\bar{M}_1$–$\bar{M}_2$ and $\bar{M}_1$–$\bar{M}_3$ pairs, obtained by separate measurements. The structure and overall magnitude of the spectral weight at $\bar{M}_2$ and $\bar{M}_3$ were benchmarked against $\bar{M}_1$ over a comparable temperature range (Fig. 3a-d). At higher temperatures, the folded feature at $\bar{M}_2$ is visibly sharper and more intense than that at $\bar{M}_3$. To quantify the transitions, we tracked the intensities of the upper and lower folded branches as functions of temperature (Fig. 3e, f). While both branches at $\bar{M}_1$ increase smoothly upon cooling, the $\overline{\mathrm{M}}_2$ and $\overline{\mathrm{M}}_3$ intensities exhibit a change in slope, with anomalies at $T^* = 160$ K and $T^{**} = 150$ K, respectively. The temperature dependence and the change in slope were reproduced for the $\overline{\mathrm{M}}_1$–$\overline{\mathrm{M}}_2$ pair on a different sample (Supplementary Note 2), confirming the robustness of the observed transitions.

To understand theoretically the observed sequential stabilization of the CDW components, we adopt a standard Landau–Ginzburg description of 1$T$-$TiSe_2$ in terms of a charge-density modulation, $\rho(\mathbf{r}) = \mathrm{Re}\{\psi_1(\mathbf{r})+\psi_2(\mathbf{r})+\psi_3(\mathbf{r})\}$, where each component is $\psi_j(\boldsymbol{r}) = A_j\, e^{i(\mathbf{q}_j \cdot \mathbf{r} + \varphi_j)}$. The wave vectors $\mathbf{q}_j$ are the CDW propagation vectors.[40,41] The Landau–Ginzburg free energy takes the form

$$F = \int \mathrm{d}^3r\ [a(\mathbf{r})\,\rho(\mathbf{r})^2 + b(\mathbf{r})\,\rho(\mathbf{r})^3 + c(\mathbf{r})\,\rho(\mathbf{r})^4 + d_0(|\psi_1\psi_2|^2 + |\psi_2\psi_3|^2 + |\psi_3\psi_1|^2)].$$

Here, the spatial dependence of $a(\mathbf{r})$, $b(\mathbf{r})$, and $c(\mathbf{r})$ encodes both the underlying lattice periodicity and the internal structure of the unit cell.[40,41] The sign of the intercomponent coupling $d_0$ determines whether a triple-$q$ or single-$q$ CDW condenses at $T_{\mathrm{CDW}}$.[40] Under the conventional assumption of sufficiently negative $d_0 < 0$ (attractive coupling), the transition favours a triple-$q$ state that emerges in a single step. This triple-$q$ CDW state subsequently undergoes a chiral transition at a lower temperature $T^*$, achieved through the development of non-zero relative phases between the three components (Fig. 4a).[42-44]

However, our measurements are consistent with repulsive coupling among the three CDW modes, i.e., $d_0 > -3/4(c_0 + c_2)$, which yields a distinct temperature-dependent sequence of phases (Fig. 4a and Supplementary Fig. 5), as detailed in Supplementary Note 4. At the highest temperatures ($T > T_{\mathrm{CDW}}$), the system retains the full $P\bar{3}m1$ symmetry without any CDW. On cooling to $T_{\mathrm{CDW}}$, a single-$q$ CDW state condenses first, represented as ($a$00), contrary to the commonly assumed triple-$q$ order. The residual symmetry of this single-$q$ state is the chiral space group $P2$ (Supplementary Note 4), arising from the concomitant lattice displacement that accompanies the charge modulation.[33] Upon further cooling to $T^*$, a secondary transition occurs in which one of the remaining two components becomes finite, producing the (*ab*0) phase, characterized by the trivial, chiral space group *P*1 (Supplementary Note 4). Depending on the parameters of the free energy, the remaining two components may turn on simultaneously, yielding the (*abc*) phase; this does not alter the symmetry of the ensuing state. In either scenario, since the (*abc*) phase is symmetry-equivalent to the (*ab*0) phase, all three components are expected to become non-zero at the lowest temperatures.

The relatively repulsive intercomponent coupling in the free energy merits further discussion. A natural microscopic origin is the competition among the three CDW components for the same low-energy electronic states. In 1*T*-$TiSe_2$, the ordering vectors $q_1$, $q_2$, and $q_3$ all hybridize the same Γ–centred Se 4*p* hole states with the three symmetry-related Ti 3*d* electron pockets at the M points. Consequently, condensation of one component reconstructs the shared Γ–centred electronic states, reducing the spectral weight available to the remaining channels and thereby generating a repulsive quartic coupling. This mechanism contrasts with the local molecular-orbital texture scenario proposed for 1*T*-$TaS_2$ and $IrTe_2$[45,46], where the dominant energy gain arises from forming real-space molecular orbitals with well-defined local symmetry. In 1*T*-$TiSe_2$, by contrast, the dominant energetic competition occurs in momentum space, where different ordering vectors compete for the same interband hybridization channels. Notably, this electronic competition coexists with cooperative lattice effects that favour simultaneous triple-*q* ordering, making the effective intercomponent coupling an emergent quantity determined by the balance between electronic competition and lattice-mediated locking, rather than a parameter fixed *a priori*. The observed sequential onset of CDW order at $T_{CDW}$ and $T^*$, separated by only a modest temperature interval, suggests that pristine 1*T*-$TiSe_2$ lies in this intermediate regime: the net coupling is sufficiently repulsive to stabilize an initial single-*q* state, yet weak enough to permit condensation of the remaining components upon further cooling. Consistent with this picture, our first-principles calculations find the competing

CDW configurations to be nearly degenerate, while the anharmonic vibrational free energy stabilizes the single-$q$ state near $T_{CDW}$ and reproduces the observed sequence of phase transitions (Supplementary Note 5).

Fig. 4b presents simulated ARPES spectra for the phases considered, which qualitatively reproduce the observed temperature evolution of the spectral weight at the different $\bar{M}$ points. In the ($a$00) phase, only $\bar{M}_1$ shows well-defined folded bands, whereas $\bar{M}_2$ and $\bar{M}_3$ remain featureless. Upon entering ($ab$0), a well-defined feature develops at $\bar{M}_2$. In the low-temperature ($abc$) phase, all three $\bar{M}$ points display well-defined folded bands. Notably, our analysis indicates that the three CDW components need not have equal amplitudes even in the ($abc$) phase, leading to different CDW gap sizes and, consequently, intensity anisotropy. Fig. 4c compares simulated ARPES spectra for an equal-amplitude ($aaa$) state, which is the non-chiral triple-$q$ state, and for an unequal-amplitude ($abc$) state with the experiment. In the trivial ($aaa$) case, dispersions and intensities are essentially indistinguishable, as verified by images and momentum-direction intensity profiles at binding energies of 0, 0.18, and 0.36 eV below $E_F$. In the ($abc$) case, the CDW gap at $\bar{M}_3$ is smaller—associated with the predominant $\mathbf{q}_3$ component—resulting in a stronger conduction-band spectral weight at the Fermi level. In addition, the valence-band spectral weights at $\bar{M}_1$ and $\bar{M}_3$ become anisotropic, with $\bar{M}_1$ stronger. These anisotropies are all reproduced in the data, confirming unequal CDW amplitudes in the fully stabilized phase, i.e., a lifted degeneracy among the three CDW components and a $P$1 ground-state symmetry.

This striking observation of the sequential emergence of the three CDW components in 1$T$-$TiSe_2$ exposes a hidden facet of the CDW transition in 1$T$-$TiSe_2$, challenging the prevailing picture established since the 1970s that a triple-$q$ state forms right at $T_{CDW}$. However, the hierarchy we uncovered is not necessarily inconsistent with accumulated measurements from the 1970s. Conventional tools for characterizing CDW–scattering methods such as x-ray

diffraction–average over domains in both the lateral and out-of-plane directions, rendering the summed signal from three single-$q$ CDW domains indistinguishable from triple-$q$.

Indeed, our calculations of the structure factor in Supplementary Fig. 9 show that averaging over three differently oriented single-$q$ domains closely reproduces the triple-$q$ pattern (Supplementary Note 6), which also rationalizes the absence of detectable chirality in scattering experiments.[34,47,48] In contrast, ARPES is surface sensitive as a result of the finite escape depth of the photoelectrons. In addition, the spatially focused probe light provides lateral resolving capability, which is demonstrated to be sufficient for the present study (Supplementary Note 3), enabling access to the inequivalent unidirectional components and the associated symmetry reduction that bulk probes average out.

Furthermore, the exposed hierarchy reconciles puzzles surrounding several recent observations. First, our identification of a chiral space group below $T_{CDW}$ is consistent with Raman spectroscopy,[33] although that study did not track the temperature evolution of individual CDW amplitudes. Second, our model explains the absence of the circular photogalvanic effect (CPGE) for $T^* < T < T_{CDW}$ and its appearance only below $T^*$, previously interpreted as a transition from a non-chiral to a chiral state. Because a finite CPGE requires at least two non-zero CDW modulations,[32] the response remains negligibly small in the intermediate regime, where only one CDW component is non-zero, even though the symmetry is already chiral ($P2$) (Supplementary Note 4). Upon cooling below $T^*$, a second component condenses and a finite CPGE can emerge. Finally, scanning tunnelling microscopy (STM), typically performed well below $T^*$, frequently reports unequal amplitudes among the three components,[30,41,49] consistent with a scenario in which the modes emerge at different temperatures and remain generically inequivalent at low temperature.

These results call for a broader perspective on CDWs—one that treats them as interacting multicomponent orders. A natural question is whether the effective repulsive intercomponent coupling identified here represents a general organizing principle for symmetry breaking in triple-$q$ CDW systems. Candidate materials include the kagome metals $AV_3Sb_5$ (A = K, Rb, Cs)[50] and the rare-earth tritellurides $RTe_3$[51], where competition between multiple CDW components is also believed to be important. Exploring the interplay between electronic competition and lattice cooperation may reveal a common physical mechanism underlying symmetry breaking across a broad class of multi-$q$ charge-density-wave materials. Another central question is how the phase sequence in 1$T$-$TiSe_2$ evolves as the overall CDW strength is weakened by external parameters, and how this multicomponent character relates to the superconductivity emerging in the vicinity of the quantum critical point of the CDW. From this perspective, our findings reframe how CDWs and related phenomena are understood and studied.

# Methods

### Sample growth and characterization

1$T$-$TiSe_2$ bulk single crystals were prepared by chemical vapour transport. Stoichiometric amounts of Ti and Se were sealed with the iodine transport agent inside an evacuated quartz tube and kept under a temperature gradient between 570 and 660 °C for 360 h. The high quality of the single-crystal samples obtained was confirmed by temperature-dependent resistivity measurements, which showed a sharp peak in the second derivative of the resistivity curve at 202.2 K as shown in Supplementary Fig. 10, indicative of the CDW transition (Supplementary Note 7).

### ARPES measurements

ARPES measurements presented in the main text were performed at beamline 7.0.2 (MAESTRO) of the Advanced Light Source (ALS) at Lawrence Berkeley National Laboratory. The measurements were performed with $p$-polarized photons with a photon energy of 111 eV which roughly corresponds to the midpoint between Γ and A high symmetry points in the bulk

Brillouin zone of the high-temperature phase. The energy resolution was better than 30 meV. The samples were cleaved *in situ* in an ultra-high vacuum chamber with a base pressure better than $5 \times 10^{-11}$ torr and aligned carefully using electrostatic deflectors to observe the orientation of the electronic structure, ensuring that the polarization vector of the incident photons is contained in the mirror plane. Data were normalized to the incident photon flux monitored by a gold mesh. The beam spot size for the measurements shown in Fig. 1 and Fig. 3c, d was approximately 34 μm × 38 μm (H × V). The beam spot size for the measurements shown in Fig. 2 and Fig. 3a, b was approximately 78 μm × 49 μm (H × V). The beam spot size estimates are based on the beamline specifications and the configuration used during each measurement. All ARPES data analysis was performed using the Python package *erlab*.[52] The spectral weight of the folded bands was obtained from energy distribution curves by integrating the intensity within a 40 meV window around the peak position as illustrated in Supplementary Note 8 and Supplementary Fig. 11.

**ARPES simulation**

For comparison with ARPES data, we performed ARPES intensity simulations based on a monolayer tight-binding model,[26] extended to include two additional transfer integrals between *p* orbitals of Se atoms across the van der Waals gap to account for the three-dimensional nature of the bulk CDW phase. The additional terms, $V_{pp\sigma}$ and $V_{pp\pi}$, were set to 0.673 eV and −0.093 eV, respectively, to reproduce the experimentally observed out-of-plane band dispersion. Additionally, the on-site energies of the Se 4*p* orbitals were adjusted to match the experimentally observed band positions. Spin-orbit coupling was included in the model, with the atomic spin-orbit coupling strength set to 0.018 eV for Ti and 0.247 eV for Se based on previous calculations.[53] A 2 × 2 × 2 supercell was used to simulate the CDW phase, with each hopping term scaled according to an empirical scaling law that describes the distance dependence of the hopping integral.[54,55] The ARPES intensity simulations were performed with the *chinook* package[56] under the three-step model of photoemission. To account for the finite probing depth of ARPES, each matrix element was scaled by an exponential factor dependent on the depth of the initial state from the surface with an inelastic mean free path λ set to 6 Å. To incorporate the spectral broadening in the $k_z$ direction, the ARPES intensity was calculated over a three-dimensional momentum grid and convolved along $k_z$ with a Gaussian distribution with the full width at half maximum set to 1 / λ.

**First-principles and anharmonic phonon calculations**

Density functional theory (DFT) calculations were performed with *Quantum ESPRESSO*[57] using the optB88-vdW functional[58] and ONCVPSP norm-conserving pseudopotentials[59], with a 110 Ry plane-wave cutoff and a 12 × 12 × 6 *k*-mesh for the 2 × 2 × 2 supercell. The four CDW configurations were relaxed in fixed 2 × 2 × 2 supercells starting from the L-point phonon eigenvectors of the pristine cell. Anharmonic phonons and vibrational free energies were evaluated within the variational self-consistent harmonic approximation[60]. Full details are given in Supplementary Note 5.

**Data availability**

The authors declare that all data required to interpret and verify the results of this study are included in the article and its Supplementary Information. All raw data are available from the corresponding authors upon request.

## Funding

This work was supported by the National Research Foundation of Korea (NRF) (grant nos. RS-2022-00143178, RS-2022-NR066723, and RS-2024-00345856) and the Korea Research Institute of Standards and Science (KRISS) (grant no. KRISS-GP2025-0015). D. H. and G. Y. C. acknowledge support for this work from the Samsung Science and Technology Foundation (project no. SSTF-BA2401-03); the NRF, funded by the Korean government through the Ministry of Science and ICT (MSIT) (grant nos. RS-2026-25479545, RS-2024-00410027, RS-2024-00444725, RS-2023-00256050, RS-2025-25453111, RS-2026-25467936 and RS-2025-08542968); the Air Force Office of Scientific Research (award no. FA23862514026); and the Institute for Basic Science (IBS) (project code IBS-R014-D1). D. H. acknowledges support for this work from the Agency for Defense Development through a grant funded by the Defense Acquisition Program Administration (DAPA) (UI257011TE). D. K. and Y. W. C. acknowledge support for this work from the NRF (grant no. RS-2024-00441954). Supercomputing resources were provided by the KISTI National Supercomputing Center (KSC-2025-CRE-0432). This research used resources of the Advanced Light Source, which is a DOE Office of Science User Facility under contract no. DE-AC02-05CH11231.

## Author contributions

Y. K. conceived the project. K. H. H. synthesized and characterized the samples. ARPES data were measured by K. H. H., Y. A., J. C., S. G., G. L., M. J., and C.-y. L., and K. H. H. and Y. K. analysed the measured data. S. H. R., C. J., A. B., E. R., and J. D. D. assisted with the ARPES experiments at ALS. M. H. and D. L. assisted with the ARPES experiments at SSRL. D. H. and G. Y. C. developed the theoretical model and performed the free energy calculations and symmetry analysis. D. K. and Y. W. C. performed the first principles and anharmonic phonon calculations. K. H. H. performed the ARPES simulations. K. H. H., D. H., G. Y. C., and Y. K. wrote the manuscript with input from all authors.

## Competing interests

The authors declare no competing interests.

# Figures

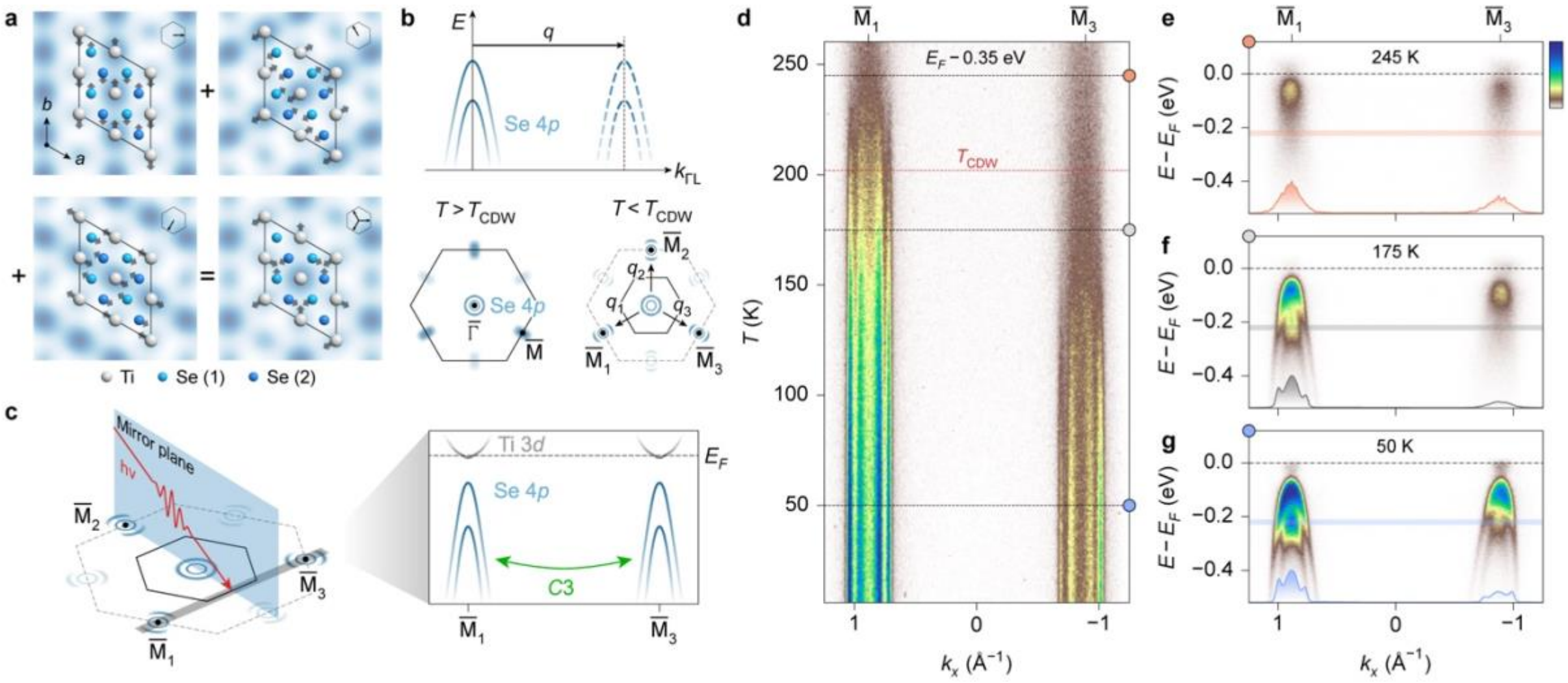

**Fig. 1 │ Overview of the chiral CDW phase in 1*T*-TiSe$_2$. a**, Displacement of ions overlaid on the corresponding charge density of each component of the triple-$q$ charge density wave (CDW) phase in 1*T*-TiSe$_2$. The displacements of the ions are not to scale. **b**, Schematic of band folding in the triple-$q$ CDW phase in 1*T*-TiSe$_2$ (top) and the electronic structure and Brillouin zone across the triple-$q$ CDW transition showing the Se 4$p$ hole bands (bottom). **c**, Experimental geometry of the angle-resolved photoemission spectroscopy (ARPES) measurement showing the mirror plane defined by the light incidence plane and one of the mirror planes of the system containing the $\overline{\Gamma}$-$\overline{\mathrm{M}}_2$ high symmetry line. Inset is the schematic band structure along the $\overline{\mathrm{M}}_1$–$\overline{\mathrm{M}}_3$ direction. **d**, Temperature-dependent electronic structure along $\overline{\mathrm{M}}_1$ and $\overline{\mathrm{M}}_3$ directions, integrated within a 20 meV window around $E = E_F - 0.35$ eV. **e–g**, Temperature-dependent energy-momentum cuts along $\overline{\mathrm{M}}_1$–$\overline{\mathrm{M}}_3$, showing the distinct onset temperatures of the CDW order at different momenta. Inset curves are the momentum distribution curves (MDCs) integrated over the energy window indicated by the shaded rectangles.

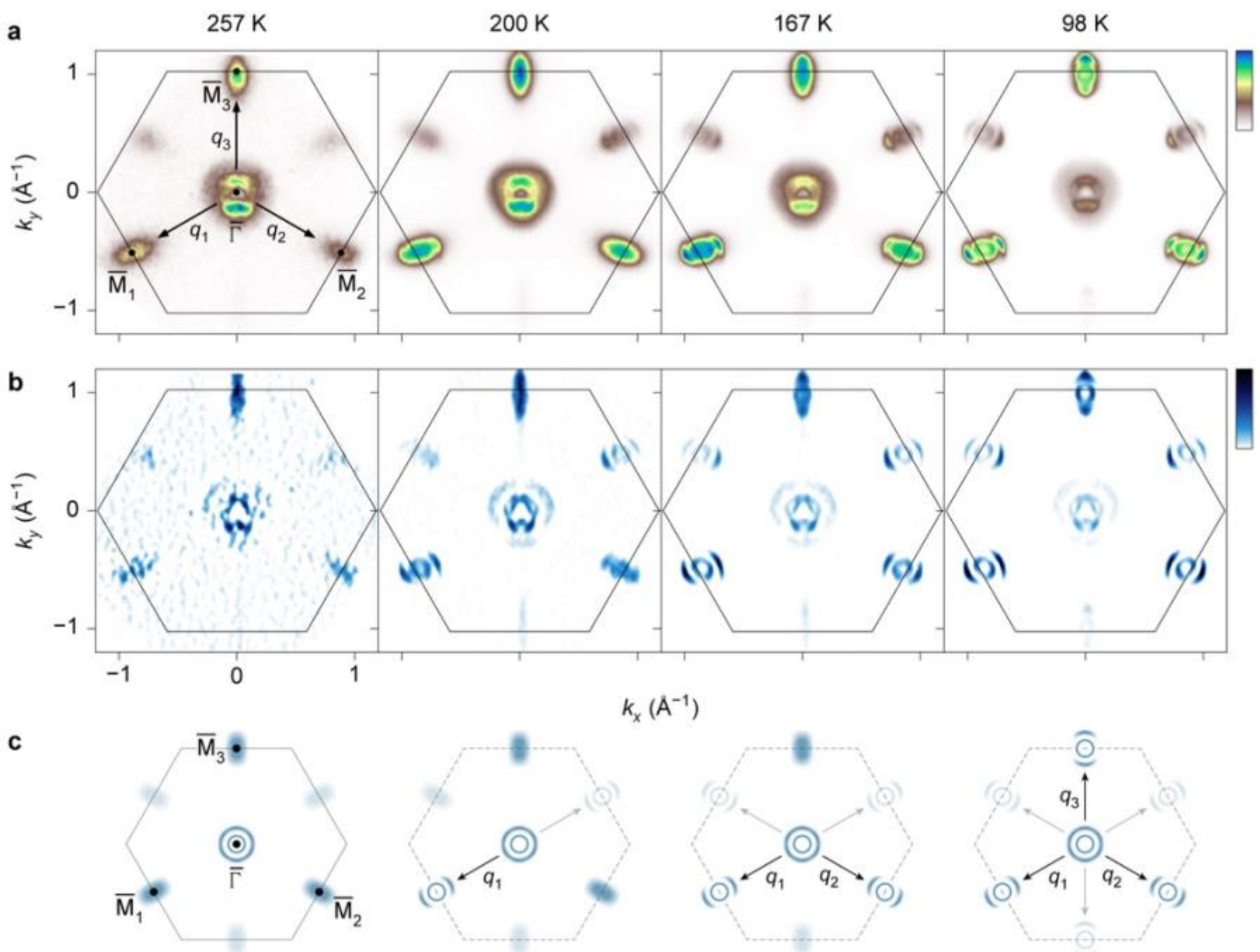

**Fig. 2 | Temperature-dependent constant-energy maps. a**, Constant-energy maps integrated within a 20 meV window around $E = E_F - 0.35$ eV. **b**, Curvature plot of the constant-energy maps in **a**. **c**, Schematic of the folded bands for different temperatures.

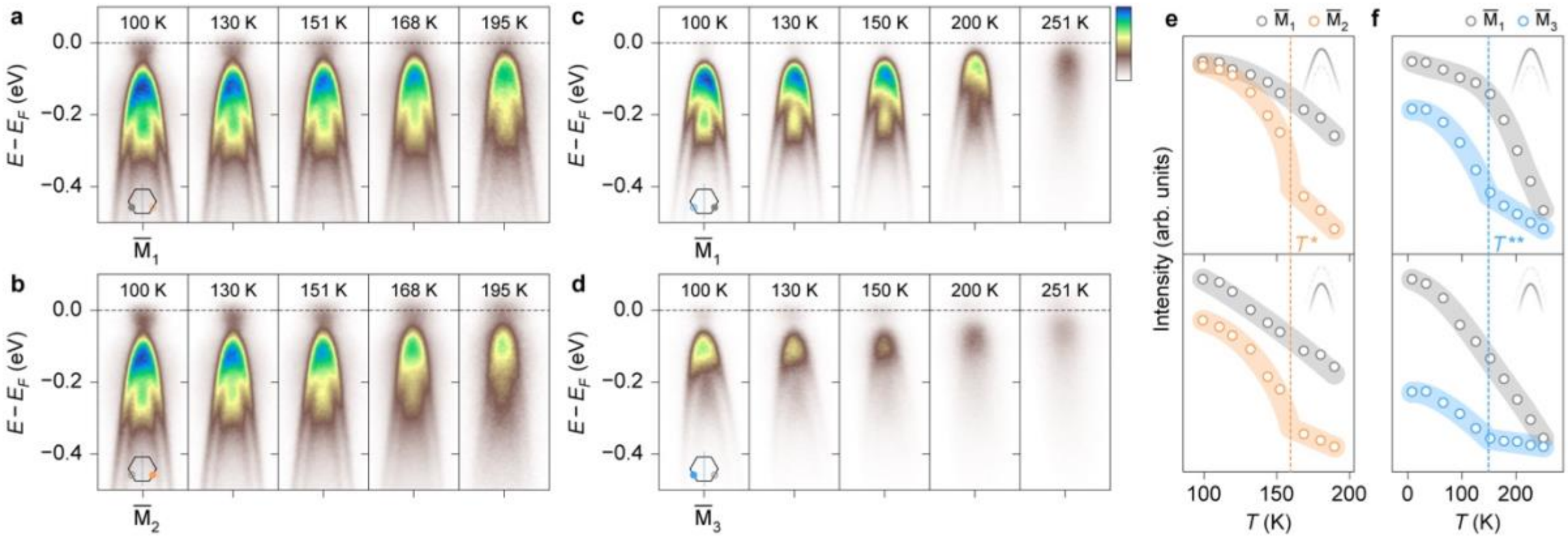

**Fig. 3 | Temperature-dependent energy-momentum cuts.** **a**, Energy-momentum cuts centred at $\bar{M}_1$ point obtained along $\bar{M}_1$–$\bar{M}_2$. Inset marks the location of the measured $\bar{M}$ point on the high-temperature Brillouin zone, with the mirror plane of the experimental geometry marked with a blue line. **b**, Same as **a**, but near $\bar{M}_2$. **c**, Energy-momentum cuts near the $\bar{M}_1$ point obtained along $\bar{M}_1$–$\bar{M}_3$. **d**, Same as **c**, but near $\bar{M}_3$. **e**, Intensity near the upper band (top) and the lower band (bottom) at $\bar{M}_1$ and $\bar{M}_2$ as a function of temperature. Shaded curves are a guide to the eye. **f**, Same as **e**, but for $\bar{M}_1$ and $\bar{M}_3$. Estimated uncertainties are smaller than the marker size, and do not affect the identification of the characteristic temperatures.

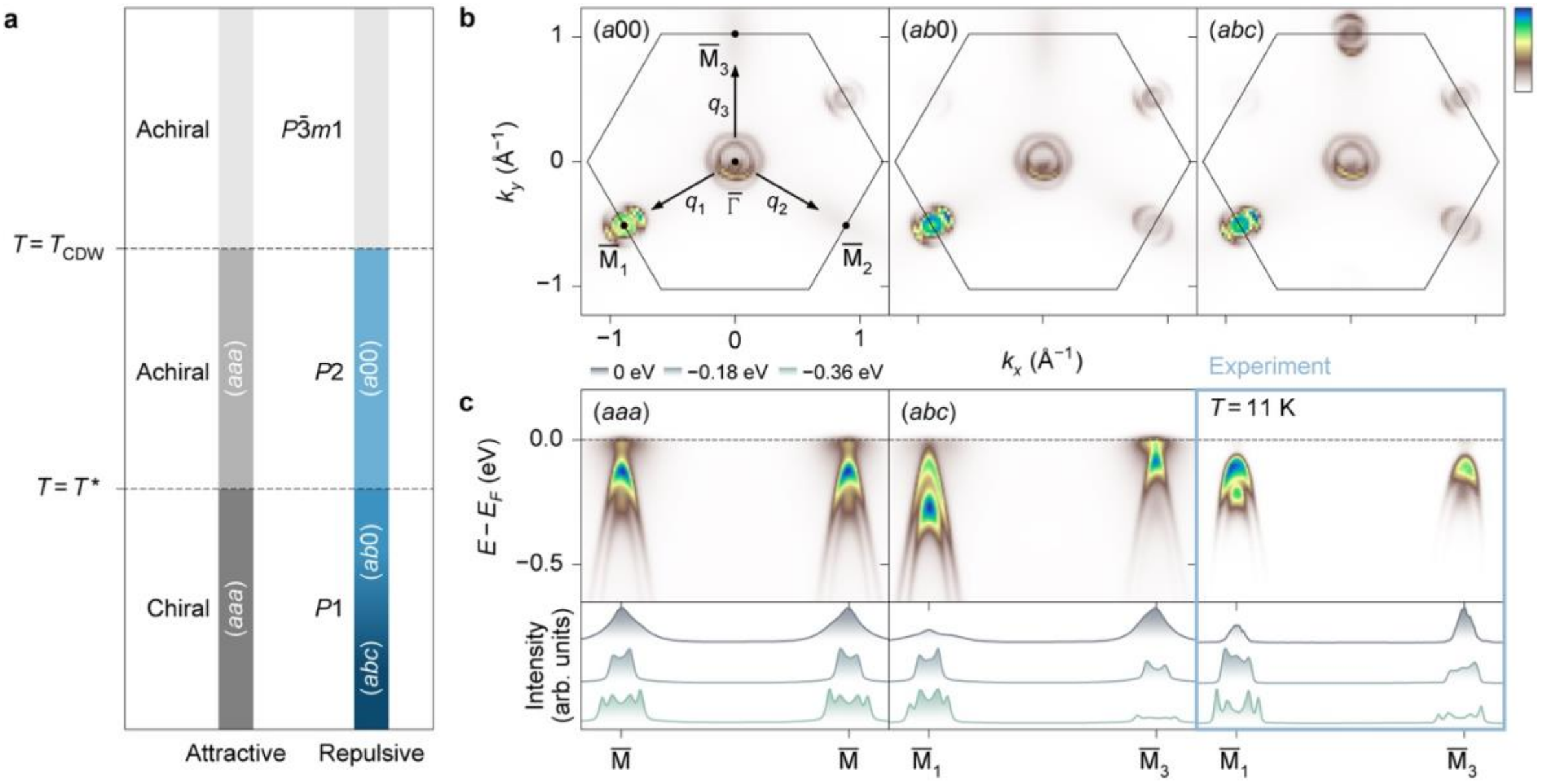

**Fig. 4 | Phase diagram and simulated angle-resolved photoemission spectra. a**, Phase evolution shown for two different cases, with effective attractive or repulsive interactions between the charge density wave (CDW) modes. In the conventional view where the three CDW modes interact attractively, the phase evolution occurs within a triple-$q$ state with equivalent amplitudes (*aaa*), evolving from an achiral to a chiral phase as the temperature decreases. On the other hand, in the case where the interactions between the modes are effectively repulsive, the phase evolution proceeds via sequential emergence of distinct wave vectors—from (*a*00) to (*ab*0)—and ultimately leads to inequivalent CDW components (Supplementary Note 4). **b**, Simulated constant-energy photoemission intensity for the (*a*00), (*ab*0), and (*abc*) phases at $E = E_F - 0.35$ eV, showing the temperature evolution of the spectral weight at the $\overline{\mathrm{M}}$ points. **c**, Simulated and experimental energy-momentum cuts along $\overline{\mathrm{M}}_1$–$\overline{\mathrm{M}}_3$. The simulated spectra from the (*aaa*) and (*abc*) phases are shown along with the experimental data taken at $T = 11$ K in the configuration shown in Fig. 1c. Insets below show the momentum distribution curves along different binding energies for each energy-momentum cut.